\documentclass[a4paper,10pt]{article}  

\usepackage{graphicx}

\usepackage{amsmath}
\usepackage{amssymb}
 
\newcommand{\EXP}[1]{\mathrm{e}^{#1}} 

\newcommand{\DEFt}{\smash{\overset{\text{\tiny def}}{=}}}
\newcommand{\imat}{{\mathrm{i}}}

\newcommand{\kete}[1]{|\kern.3ex#1\kern.3ex\rangle}

\newcommand{\brae}[1]{\langle\kern.3ex #1 \kern.3ex|} 
\begin{document}

\title{A mechanical model of tunnelling}
\author{Amaury Mouchet\\ \\
Laboratoire de Math\'ematiques 
  et Physique Th\'eorique,\\ Universit\'e Fran\c{c}ois Rabelais de Tours 
--- \textsc{\textsc{cnrs (umr 6083)}},\\
F\'ed\'eration Denis Poisson,\\
 Parc de Grandmont 37200
  Tours,  France. \\{mouchet@phys.univ-tours.fr}}

\date{\today}

\maketitle 


\begin{abstract}It is shown how the model which was introduced in \cite{Mouchet08a} allows to mimic the quantum tunnelling between two symmetric
one-dimensional wells.
\end{abstract}

Following the publication of \cite{Mouchet08a}, some readers asked me
to provide more details on the concluding remark about tunnelling
(first paragraph of \S~6).  Strictly speaking, tunnelling is a quantum
phenomenon that refers to any process that is classically forbidden
(\textit{i.e.} cannot be understood from the \textit{real} solutions
of the Hamilton equations).  The paradigmatic example (that can be
found in many textbooks, like \cite{Messiah65a}) is given by the oscillations
of a quantum system between two one-dimensional potential wells
separated by a barrier whose maximal energy is larger than the energy
of the system.  As understood as early as 1927 by Hund
\cite{Merzbacher02a}, it relies on the existence of some evanescent
Schr\"odinger waves connecting the two wells. Therefore, it is
perfectly justified to consider a purely classical mechanical model to
illustrate tunnelling provided it exhibits the appropriate evanescent
waves (we do not intend here to compare the classical model with its
quantum analogue). In optics, where evanescent waves can be easily
created with dielectric materials, the analogue of quantum tunnelling
has already been considered, for instance in \cite{Nockel/Stone97a}.

To mimic the double-well symmetric situation, let us consider two
identical oscillators coupled to the Klein-Gordon string (as described
in \cite{Mouchet08a}) and located at $x=\pm a/2$ ($a>0$). When the
oscillation frequency of the free oscillators~$\Omega_0$ is below the
``infra-red'' cut-off on the Klein-Gordon string~$\omega_0$, there are
two stable modes, one symmetric and one antisymmetric whose
corresponding frequencies are denoted by~$\omega_+$ and $\omega_-$
respectively : If the two oscillators were infinitely far away, we
would get~$\omega_+=\omega_-=\omega_b$ given by (42) of
\cite{Mouchet08a} but at finite distance the string connecting the two
oscillators lifts the degeneracies between the two frequencies : the
splitting $\Delta\omega\DEFt\omega_--\omega_+\neq0$ represents the
frequency of the beating between the two oscillators.  At these
frequencies~$\omega_\pm$, both below~$\omega_0$, no energy transport
along the string is allowed on the average (equation (9) of
\cite{Mouchet08a}) and however some energy can be exchanged between
the two oscillators.

This is similar with the quantum situation: for the quantum symmetric
double-well potential, the true eigenstates are delocalised between
the two wells with energies~$E_+$ and~$E_-$; a state that is localised
in one well only is a linear superposition of a symmetric and an
antisymmetric state and oscillate back and forth between the two wells
at a frequency~$(E_--E_+)/\hbar.$

The tunnelling splitting is expected to be exponentially small with
the distance between the two oscillators \cite{Garg00a} as can be
proved by the following computation:

For~$x<-a/2$, we keep an evanescent wave on the string with the form
$D_-\EXP{\imat(-k(x+a/2)-\omega t)}$ with
$k=\imat\sqrt{\omega_0^2-\omega^2}$ whereas for ~$-a/2<x<0$ we have
$C_+\EXP{\imat(k(x+a/2)-\omega t)}+D_+\EXP{\imat(-k(x+a/2)-\omega t)}$.
 The complex coefficients $C_+$ and $D_+$ are linearly related
to~$D_-$ with the matrix~$T(\omega)$ given by (23-26) of
\cite{Mouchet08a}.  The antisymmetric mode corresponds to an odd
wavefunction that vanishes at $x=0$: $C_+\EXP{\imat
  ka/2}+D_+\EXP{-\imat ka/2}=0$.  The symmetric mode corresponds to an
even wavefunction whose spatial derivative vanishes at $x=0$ :
$C_+\EXP{\imat ka/2}-D_+\EXP{-\imat ka/2}=0$.  In both cases, a
non-trivial solution can be found if
$\rho=\pm\EXP{-\imat ka}$. The frequencies $\omega_\pm$ must be the
solutions of the two equations respectively
\begin{equation}
  1+\frac{2\sqrt{\omega_0^2-\omega_{}^2}}{\kappa}\frac{\omega^2-\Omega_\kappa^2}{\omega^2-\Omega_0^2}
=\mp\,\EXP{-a\sqrt{\omega_0^2-\omega^2}}\;.
\end{equation}
If the right hand side vanished ($a\to+\infty$ while keeping fixed all the other parameters),
 we would recover the equation that furnishes $\omega_b$ (see equation (38) of \cite{Mouchet08a}).
For finite~$a$, in the case where the right hand side is small ($a\sqrt{\omega_0^2-\Omega_0^2}\gg1$), the first order expansion in the 
coupling parameter~$\kappa$ provides
\begin{equation}
  \Delta\omega\simeq\frac{\kappa^2}{2\Omega_0\sqrt{\omega_0^2-\Omega_0^2}}\,\EXP{-a\sqrt{\omega_0^2-\Omega_0^2}}\;.
\end{equation}


\end{document}